\documentclass[lettersize,journal]{IEEEtran}
\usepackage{amsfonts}
\usepackage{amssymb}
\usepackage[ruled,vlined]{algorithm2e}
\usepackage{mathtools}  
\usepackage{amsthm}
\usepackage[]{algorithm2e}
\usepackage{subfigure}
\usepackage{tabulary}
\usepackage{footnote}
\usepackage[export]{adjustbox}
\usepackage{booktabs}
\usepackage[justification=centering]{caption}
\usepackage{comment}

\usepackage{cases}
\usepackage{cite}
\usepackage{multicol}
\usepackage{xcolor}
\usepackage{graphicx}
\graphicspath{{./}{figures/}}

\usepackage{stfloats}
\usepackage{cuted}
\usepackage{lipsum}

\usepackage{subfigure}

\title{Optimal Beam Training for mmWave Massive MIMO using 802.11ay}
\author{Lyutianyang Zhang and Sumit Roy, \IEEEmembership{Fellow, IEEE}}
\date{April 2019}
\begin{document}
 \pagenumbering{gobble}
\maketitle
\begin{abstract}
Beam training of 802.11 ad is a technology that helps accelerate the analog weighting vector (AWV) selection process under the constraint of the existing code-book for AWV. However, 5G milli-meter wave (mmWave) multiple-input-multiple-output (MIMO) system brings challenges to this new technology due to the higher order of complexity of antennae. Hence, the existing codebook of 11ad is unlikely to even include the near-optimal AWV and the data rate will degrade severely. To cope with this situation, this paper proposed a new beam training protocol combined with the state-of-the-art compressed sensing channel estimation in order to find the AWV to maximize the optimal data-rate. Simulation is implemented to show the data-rate of AWV achieved by 11 ad is worse than the proposed protocol.

\end{abstract}

\section{Introduction}
The growing demand for bandwidth is driving the adoption of millimeter-wave systems (\textbf{mmWave}) in cellular last-mile access technology. As part of the 5G revolution, mmWave is expected to support gigabit rates at millisecond latency. Such high performance will be enabled by beamforming base stations equipped with a large number of mmWave antennas. 

Beamforming is crucial for overcoming the significant increase in over-the-air path loss at mmWave bands. As is well-known, such increased channel gain is required at {\em both} ends of a link, implying both Tx and Rx must implement beamforming meeting the link budget for the desired data rates or modulation and coding scheme (MCS) employed. In turn, this introduces a different problem - of beam alignment. When the antenna radiation pattern is a narrow spot beam, the increased antenna gain is confined to a small angular region. Thus, the benefits of beamforming - net beamforming gain on both the Tx and Rx sides- is available only if the two beams are pointed in an optimal manner. \textbf{Beam training} refers to the process of finding the best beam alignment from an initial state of little or no information about the channel. In contrast, \textbf{beam tracking} refers to the process of maintaining beam alignment when the devices are moving during communication. 

The need for beam training and beam tracking comes at a price - significant resource overhead must be added to the cellular systems to achieve the needed beamforming gains in real systems. Because last-mile mmWave deployments will be implemented in small cells, user mobility implies frequent entry and departure and consequent need for frequent initial beam training. In turn, this will dominate the overhead complexity and  latency constraints that must be budgeted for. For massive MIMO mmWave systems, beam-training implies accurate channel state (CSI) whose complexity grows polynomially as a function of key system parameters. 

In summary, the great challenge in mmWave systems design is to find the sweet spot between achieving beam pointing accuracy that necessarily involves requisite CSI complexity and feedback latency, and what is achievable within the 11ad/11ay beam-training architecture that must necessarily manage complexity and latency budgets. Our approach is to start from the latter as a point of departure and seek an efficient beam training approach that achieves enhanced beam training accuracy with limited complexity/latency overhead; this is in contrast to most of the academic mmWave MIMO signal processing literature that provides optimal CSI solutions but which are largely infeasible for integration within the 11ad/11ay MAC.

\begin{figure*}[t]
  \centering
  \includegraphics[width=100mm, height=20mm]{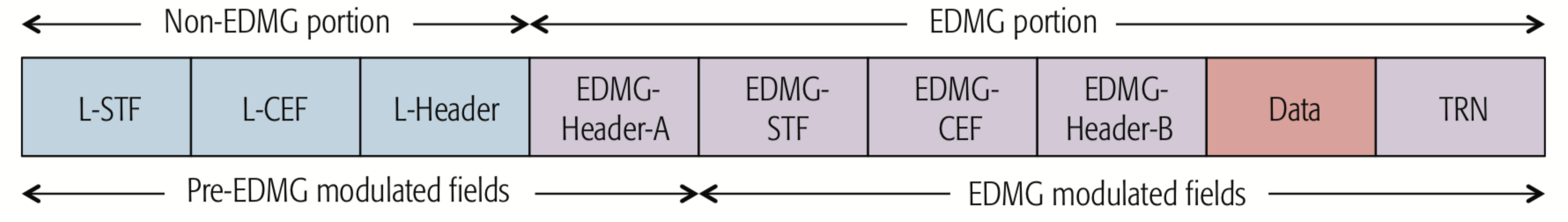}
\caption{} 
\label{fig:pro1}
\end{figure*}

\subsection{Literature Review}
mm Wave communication is a promising technology for next-generation wireless communication owing to its abundant frequency spectrum resource, which promises a much higher capacity than the exist- ing wireless local area networks (WLANs) and the current cellular mobile communication. In fact, mmWave communication has received increasing attentions as an important candidate technology in both the next-generation WLANs \cite{perahia2010ieee,xiao2013suboptimal,xia2008practical} and mobile cellular communication \cite{xia2008multi,khan2011mmwave,zhang2022architecture}. A fundamental challenge to mmWave communication is the extremely high path loss, thanks to the very high carrier frequency on the order of 30-60 GHz. To bridge this significant link budget gap, joint Tx/Rx beamforming is usually required to bring large antenna array gains, which typically requires large Tx/Rx antenna array size (e.g. array size of 36 [4]). Thanks to the small wavelength at the mmWave frequency range, large antenna arrays may be packed into a small area. Hence in our paper, we address the beamforming problem at the {\em beginning phase} of 11 ay for hybrid MIMO architecture sub-connected antenna arrays, {\em in the absence of any CSI}.

In \cite{xiao2016hierarchical}, hierarchical codebook design is introduced to decrease the total steps to reach the optimal beam in the codebook in single RF-chain MIMO architecture. However, this codebook design requires multiple feedback rounds to close the link. One the contrary, we propose an one-round beamtraining method for hybrid MIMO sub-connected architecture with multiple RF-chain on both transmitter and receiver. In \cite{gao2016energy}, the optimal precoding matrix is expressed in closed form given the hybyrid sub-connected architecture with multiple RF-chain. However, full CSI is assumed and the optimal precoding matrix is obtained that maximizes link capacity between the transmitted and the received RF signal between the antennas. Although \cite{el2017} considers the precoder and combiner as a whole in optimizing the precoding matrix, again the full CSI is assumed and the system architecture is not sub-connected but fully-connected. We propose, on the other hand, a method which obtains optimal beamforming solution without CSI and the state-of-the-art sub-connected MIMO architecture. \cite{molisch2017hybrid} studied the hybrid beamforming in hyrbrid sub-connected massive MIMO system and proves that sub-connected architecture can achieve the exactly same performance as fully-connected architecture. However, full CSI is still assumed. 

In this paper, we consider the problem of optimizing initial beam training for hybrid sub-connected MIMO RF transceiver. Beginning with no initial CSI at receiver, we explore what is feasible with one-round of algorithmic design optimization and feedback to transmitter. Note that we use beam pattern vector (BPV) and analog weighting vector (AWV) interchangeably throughout this paper. 

\emph{Notation}: Lower-case and upper-case boldface letters denote vectors and matrices respectively; $(\bullet)^{T}$, $(\bullet)^{H}$, $(\bullet)^{-1}$, and $|\bullet|$ denote the transpose, conjugate transpose, inversion, and determinant of a matrix, respectively; $||\bullet||_1$ and $||\bullet||_2$ denote the $l_1$ and $l_2$ norm of a vector, respectively. $\mathbf{I}_{N}$ denotes a $N \times N$ identity matrix.

\begin{figure*}[h]
  \centering
  \includegraphics[width=160mm, height=80mm]{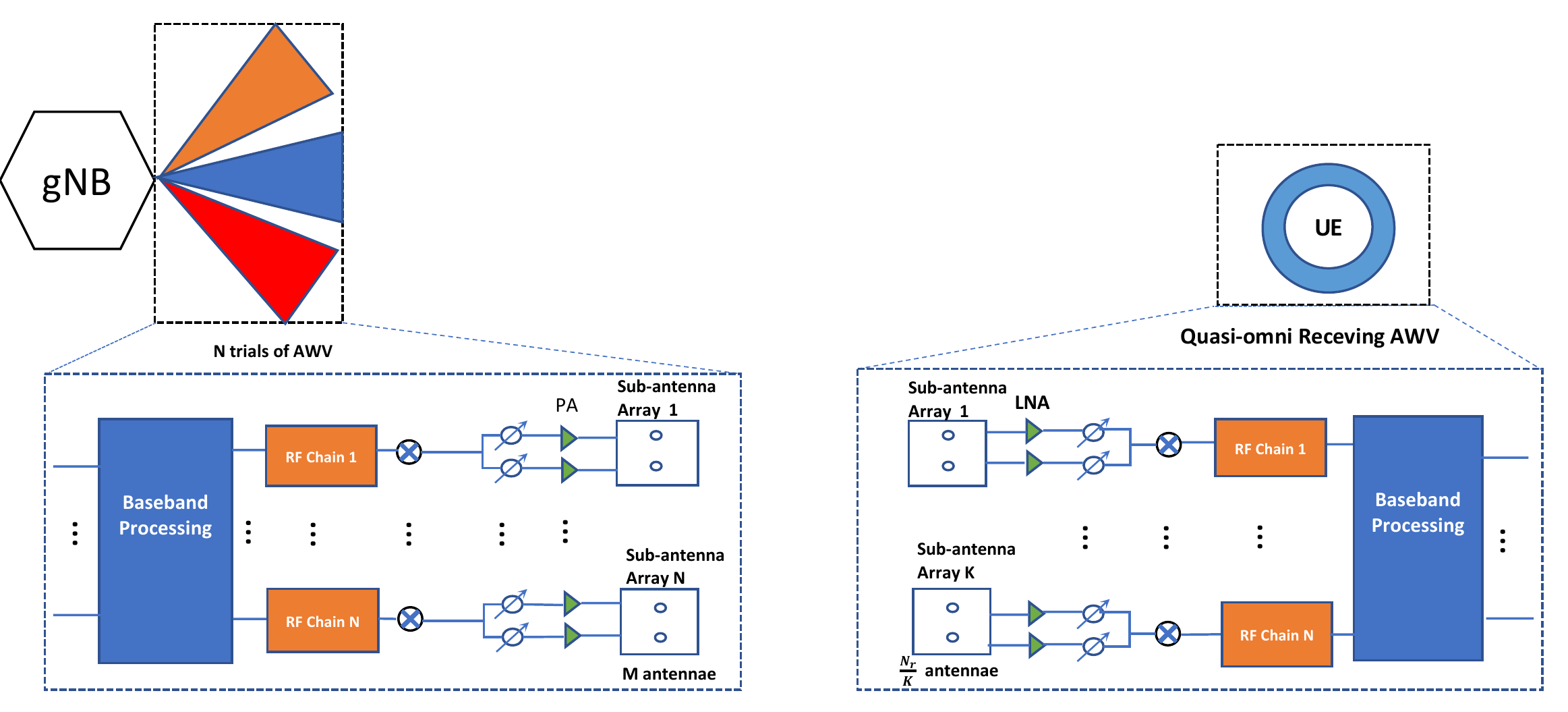}
\caption{} 
\label{fig:sysm}
\end{figure*}
\section{system model}

\subsection{Formulation for beam training in 802.11ay}
IEEE 802.11ay aims to do beam-forming with no or almost no CSI. We propose time-efficient beamforming algorithms to find the optimal AWV in one-shot, i.e., only one-time beamforming feedback from the receiver to transmitter after all AWVs has been tried. Now, we introduce the system model for massive MIMO hybrid architecture with sub-connected antenna arrays. In sub-connected architecture,  N data streams in the baseband are precoded by the digital precoder $D$. In cases where complexity is a concern, $D$ is designed to be a diagonal matrix $\text{diag}[d_1,d_2,\dots,d_{N}]$, where $d_{n} \in R$ for $n=1,\dots,N$. Then the role of D essentially performs some power allocation. After passing through the corresponding RF chain, the digital-domain signal from each RF chain is delivered to only M PSs to perform the analog precoding, which can be denoted by the analog weighting vector $\mathbf{a}_{n} \in \mathcal{C}^{M \times 1}$, whose elements have the same amplitude $\frac{1}{\sqrt{M}}$ but different phases. After the analog precoding, each data stream is finally transmitted by a sub-antenna array with only M antennas associated
with the corresponding RF chain. The receiver also has analog combiner and digital combiner and we denote $\mathbf{W}=\tilde{\mathbf{D}}\tilde{\mathbf{A}}$, where $\tilde{\mathbf{D}} \in \mathcal{C}^{K \times K}$ and $\tilde{\mathbf{A}} \in \mathcal{R}^{K \times N_r}$ denote digital combiner and analog combiner respectively. 
Then the received
signal vector $\mathbf{y}=[y_1,\dots,y_K]^{T}$ at the user in a narrow band system can be presented as 
\begin{equation}
    \begin{aligned}
    \mathbf{y}=\sqrt{\rho} \mathbf{W}\mathbf{H}\mathbf{A}\mathbf{D}\mathbf{s}+\mathbf{W}\mathbf{n}, 
    \end{aligned}
\end{equation}
where $\rho$ is the average received power. $\mathbf{H} \in \mathcal{C}^{N_r \times N_t}$ denote the channel matrix, and $N_t=M \times N$. $\mathbf{A}$ is the $NM \times N$ analog precoding matrix comprising N analog weighting vector $\{{\mathbf{a}}_m\}^{N}_{m1}$ as 
\begin{equation}
    \mathbf{A}=\begin{bmatrix}
    \mathbf{a}_1  & 0 & \dots & 0\\
    0 & \mathbf{a}_2 & & 0\\
    \vdots&& \ddots& \dots\\
    0 & 0 & \dots & \mathbf{a}_N
    \end{bmatrix}_{NM \times N },
\end{equation}
where $\mathbf{s}=[s_1,s_2,\dots,s_N]^T$ represents the transmitted signal vector in the base band. In this paper, we assume the widely used Gaussian signals with normalized signal power $\mathbf{E}[ss^{\dagger}]=\frac{1}{N}I_{N}$.$\mathbf{P}=\mathbf{A}\mathbf{D}$ presents the hybrid precoding matrix. $\mathbf{n}=[n_1,n_2,\dots,n_{N_r}]^T$ is an AWGN vector, whose entries follow the i.i.d.  $\mathcal{CN}(0,\sigma^2)$. Finally, in the transmitter beam training phase, the receiver adopts an omni-directional receiving pattern which indicate the following analog combinging matrix 

\begin{equation}\label{eq:Abar}
    \begin{aligned}
    \bar{\mathbf{A}}=\begin{bmatrix}
    \bar{\mathbf{a}}_1  & 0 & \dots & 0\\
    0 & \bar{\mathbf{a}}_2 & & 0\\
    \vdots&& \ddots& \dots\\
    0 & 0 & \dots & \bar{\mathbf{a}}_K
    \end{bmatrix}_{K \times N_r},
    \end{aligned}
\end{equation}
where the number of antennae in each sub-antenna array is $\frac{N_r}{K}$, and $ \bar{\mathbf{a}}_k \in \mathcal{C}^{1 \times \frac{N_r}{K}}$. The digital combiner matrix $\bar{\mathbf{D}} = \text{diag}[\bar{d}_1,\bar{d}_2,\dots,\bar{d}_K]$. Assume now we are in the phase of transmitter beam training phase, in which receiver is using an fixed omni-directional antenna to receive 
$N$ trials of different AWV pattern of $\mathbf{A}$ with identity digital precoding matrix $\mathbf{D}=I_{N}$.

Now we give our orthogonal beamtraining codebook design and beamtraining solution for transmitter.
\begin{algorithm}

\KwResult{$\mathbf{A}^{est}$}
~Design an orthogonal codebook for $\mathbf{A}$ in which $\mathbf{a}_i^{\dagger} \mathbf{a}_j=1$.
Then do N trial of transmitting AWV by cyclic rotation on the block diagonal matrix $\mathbf{A}$. For example, after one cyclic rotation (we define the rotation number as k=1 and k=0 when there is no rotation), the analog precoding matrix is 
\begin{equation}
\begin{bmatrix}
\mathbf{a}_N  & 0 & \dots & 0\\
0 & \mathbf{a}_1 & & 0\\
\vdots&& \ddots& \dots\\
0 & 0 & \dots & \mathbf{a}_{N-1}
\end{bmatrix}_{NM \times N }, 
\end{equation}  and so on.  We denote $\mathbf{Y}=[\mathbf{y}_1,\mathbf{y}_2,\dots,\mathbf{y}_N]_{K \times N}$ in which $\mathbf{y}_n$ is the received signal at $n^{th}$ trial. 

$i=0$\;
$\mathbf{a}^{est}_{k}=zeros(M,1)$\;
\While{$i<N$}{
\eIf{$k-i+1>0$}{
$\mathbf{a}^{est}_{k}=\mathbf{a}^{est}_{k}+\frac{\mathbf{y}_{i+1}^{\dagger}\mathbf{y}_{i+1 }}{\sum_{n=1}^{N}||\mathbf{y}_n||^{2}_{2}}\mathbf{a}_{k-i+1}$\;
}
{
$\mathbf{a}^{est}_{k}=\mathbf{a}^{est}_{k}+\frac{\mathbf{y}_{i+1}^{\dagger}\mathbf{y}_{i+1}}{\sum_{n=1}^{N}||\mathbf{y}_n||^{2}_{2}}\mathbf{a}_{k-i+1+N}$\;
}
$i=i+1$\;
}
Construct $\mathbf{A}^{est}$ as shown in Eq. \eqref{eq:Abar}.
 \caption{Orthogonal beam training algorithm for sub-connected MIMO architecture for IEE 802.11 ay}
\end{algorithm}

\section{Simulation}
\subsection{Channel model}

It is know that mmWave channel $\mathbf{H}$ is not likely to have the property of rich-scattering model assumed at low frequencies due to limited number of scatters in the mmWave propagation environment \cite{pi2011introduction}. In this paper, we adopt the geometric Saleh-Valenzuela channel model to embody the low rank and spatial correlation characteristics of mmWave communications \cite{el2017} as
\begin{equation}
    \mathbf{H}=\gamma \sum_{l=1}^{L} \alpha_{l} \Lambda_{r}(\phi^{r}_{l},\theta_{l}^{r}) \Lambda_{t}(\phi^{t}_{l},\theta_{l}^{t})\mathbf{f}_{r}(\phi_{l}^{r},\theta_{l}^{r})\mathbf{f}_{t}^{h}(\phi_{l}^{t},\theta_{l}^{t}), 
\end{equation}
where $\gamma=\sqrt{\frac{N_r N_t}{L}}$ is a normalization factor, $L$ is the number of effective channel paths corresponding to limited number of scatters, and we usually have $L \leq N$ for mmWave communication systems, $\alpha_{l} \in \mathcal{C}$ is the gain of the $l^{th}$ path. $\phi_{l}^{t}(\phi_{l}^{t})$ and $\phi_{l}^{r}(\phi_{l}^{r})$ are the azimuth (elevation) angle of departure and arrival (AoDS/AoAs), respectively, $\Lambda_{t}(\phi_{l}^{t},\theta_{l}^{t})$ and $\Lambda_{r}(\phi_{l}^{r},\theta_{l}^{r})$ denote the transmit and receive antenna array gain at a specific AoD and AoA, respectively. For simplicity but without loss of generality, $\Lambda_{t}(\phi_{l}^{t},\theta_{l}^{t})$ and $\Lambda_{r}(\phi_{l}^{r},\theta_{l}^{r})$ can be set as one within the range of AoDs/AoAs \cite{alkhateeb2013hybrid}. Finally, $\mathbf{f}_{t}(\phi_{l}^{t},\theta_{l}^{t})$ and $\mathbf{f}_{r}(\phi_{l}^{r},\theta_{l}^{r})$ are the antenna array response vectors depending on the antenna array structures at the BS and the user, respectively. We use the uniform linear array (ULA) with $U$ elements in this paper without the loss of generality, the array response vector (ARV) can be presented as 
\begin{equation}
    \mathbf{f}_{ULA}(\phi)=\frac{1}{\sqrt{U}}[1,e^{j\frac{2 \pi}{\lambda}d\sin(\phi)}, \dots,e^{j\frac(U-1){2 \pi}{\lambda}d\sin(\phi)}],
\end{equation}
where $\lambda$ denotes the wavelength of the signal, and d is the antenna spacing. 

\subsection{Performance analysis of COM and 11 ay}
Firstly, we construct a codebook design according to the requirement of orthogonality, i.e., $\mathbf{A}^{\dagger}\mathbf{A}=I$, which is also equivalent to $\mathbf{a}_{i}^{\dagger}\mathbf{a}_{j}=1$ unless $i=j$ and $0$ otherwise. Channel matrix is assumed to be unknown through entire beamforming phase and $E[||H||^{2}_F]=N_{t} N_{r}$. Signal power $E[\mathbf{s}^{\dagger}\mathbf{s}]=\frac{1}{N}\mathbf{I}_{N}$. AWGN vector, whose entries follow the i.i.d. $\mathcal{CN}(0,\sigma^2)$. $f=28GHz$ and $\lambda=\frac{c}{f}$, where $c=3*10^8 m/s$. $d=\lambda/2$. AoDs are assumed to follow the uniform distribution within $[-\pi/6,\pi/6]$.AoA follows uniform distribution within $[-\pi,\pi]$. $L=3$. In the beginning phase of beamforming, receiver does not have information of CSI, so we assign equal power to each RF chain, i.e., $\tilde{\mathbf{D}}=I$ and $\tilde{\mathbf{A}}$ is a constant matrix. $N=8$ $M=8$ $N_t=64$ $N_r=16$ $K=4$. SNR is $\frac{\rho}{\sigma^2}$.

The performance of COM is measured in terms of effective channel capacity between the transmitter's data stream and the receiver's data stream, which can be expressed as follows \cite{goldsmith2003capacity},
\begin{equation}
    C=\log \left( |I_{K}+\frac{\rho}{N }\mathbf{R}_{n}^{-1}\mathbf{W}\mathbf{H}\mathbf{P}^{H}\mathbf{H}^{H}\mathbf{W}^{H}  | \right),
\end{equation}
where $\mathbf{R}_{n}=\sigma_{n}^2 \mathbf{W}\mathbf{W}^{H}$.

\begin{figure}[t]
  \centering
  \includegraphics[width=59mm, height=25mm]{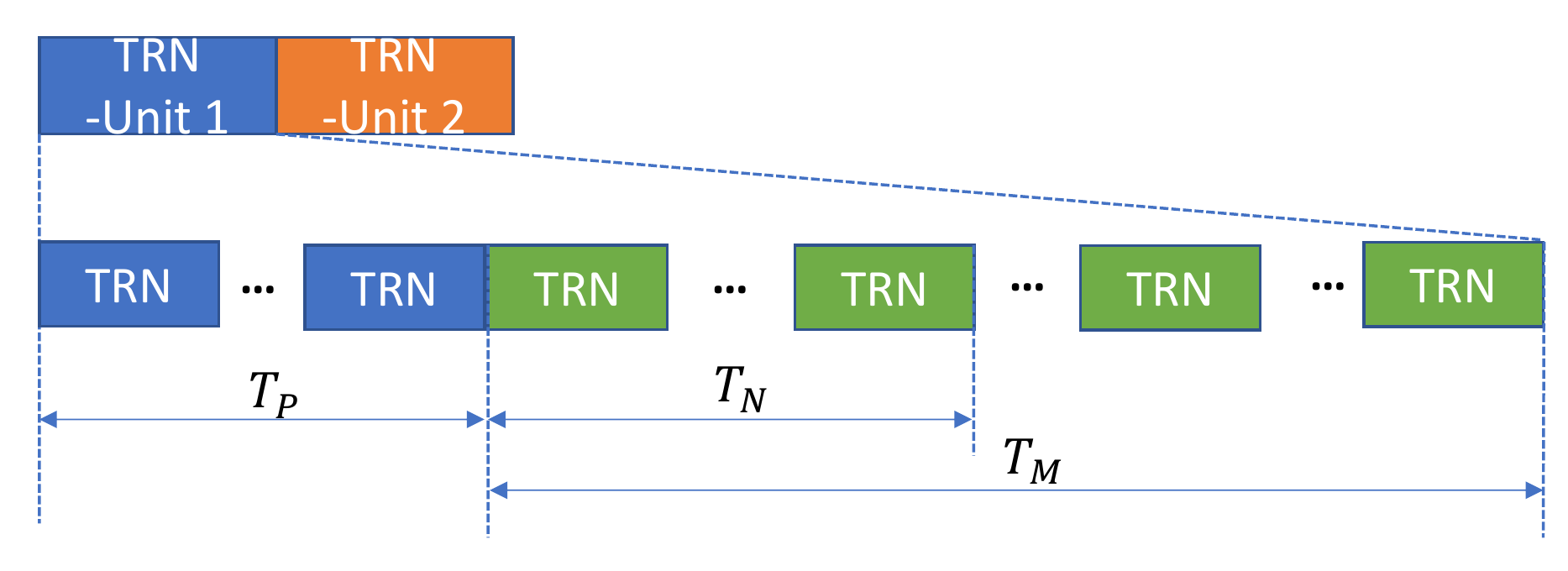}
\caption{IEEE 802.11 ay beam training protocol for receiver \cite{ghasempour2017ieee}} 
\label{fig:pro2}
\end{figure}
\begin{figure}[h]
  \centering
  \includegraphics[width=80mm, height=50mm]{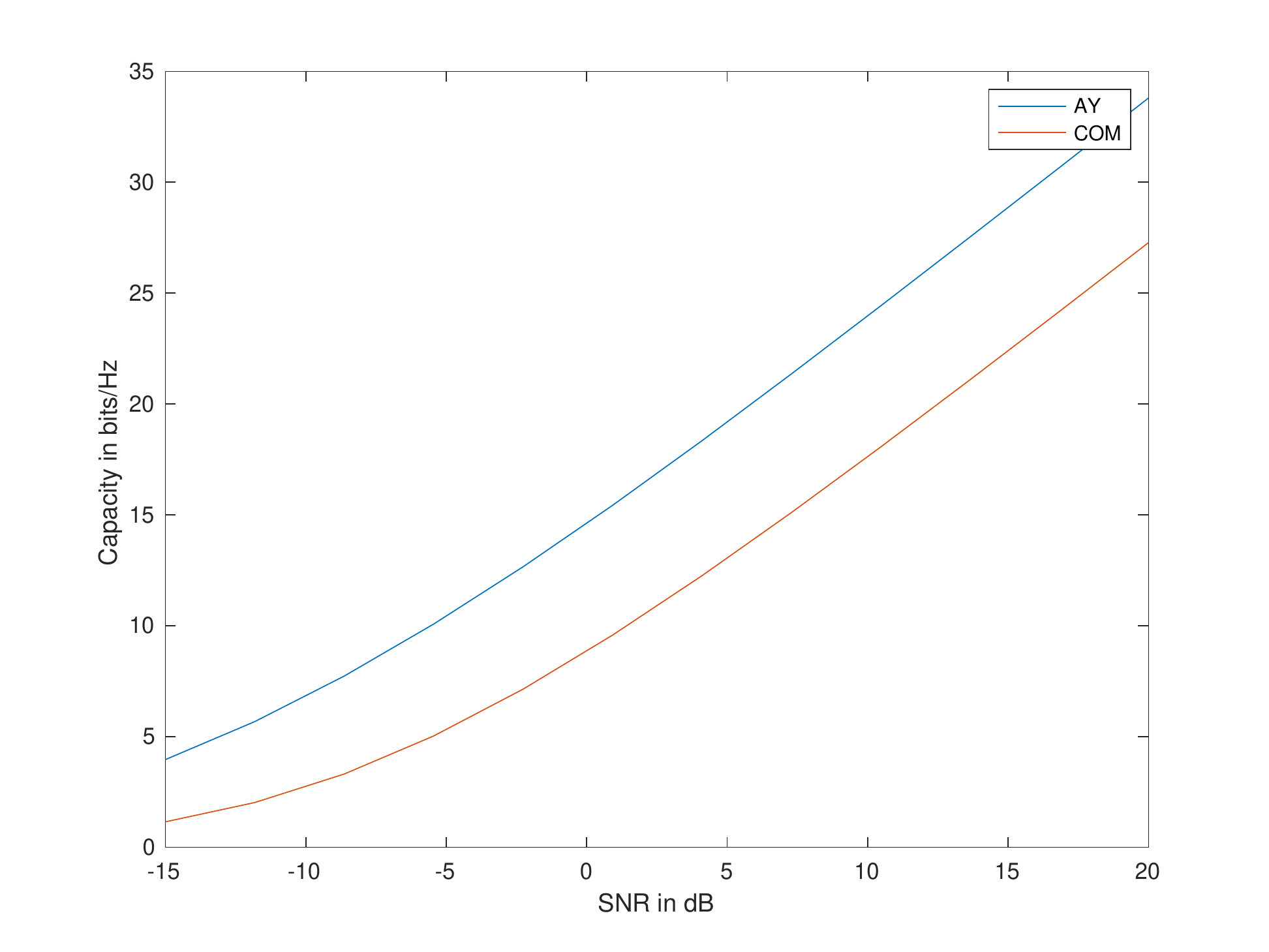}
\caption{EDMG BRP-TX packets} 
\label{fig:Sim}
\end{figure}

IEEE 802.11 ay beam training and its training (TRN) field format is shown in Fig. \ref{fig:pro2}, which introduces a beam refinement protocol (BRP) that can improve its antenna configuration for transmission. The transmitter uses difference AWVs in the transmission of the TRN field while the receiver uses the same AWV in its reception. Three parameters define the format and length of a TRN-Unit used for transmit beamforming training, also called as EDMG BRP-TX packets: EDMG TRN unit $T_P$, EDMG TRN-Unit $T_M$, and EDMG TRN-Unit $T_N$. In a TRN-Unit, the first $T_P$ TRN subfields are transmitted with the same AWV as the data field. Therefore, the receiver may use such TRN subfields to maintain synchronization, for which we do not consider in thie paper. In the transmission of the remaining $T_M$ TRN subfields of a TRN-Unit, the transmitter changes AWV at the beginning of each TRN subfield. In order to improve the robustness of the beamforming training process, of the last $T_M$ TRN subfields of a TRN-Unit, more than one consecutive TRN subfield may be transmitted with the same AWV, in this paper, we do not consider the repeated transmission of same AWV. The number of the consecutive TRN sub-fileds tranmitted with the same AWV is $T_N$. In standard IEEE 802.11 ay, all received signal of transmitter AWV are compared to yield the one with the highest energy and consequently its corresponding transmitter AWV, the, this AWV is fed back to transmitter. In this paper, however, Algorithm.1 follows the reception of all signal and its performance is shown in Fig. \ref{fig:Sim}. The comparison between COM and standard 802.11 ay shows that there is around 4 bit/Hz gain. The gain increases as the SNR increases. Moreover, the time complexity of this algorithm is only $O(n)$ and is does not do any complex matrix decomposition like SVD in most of the state-of-the-art precoding method.

\section{Conclusion}
Traditional 11 ad and 11 ay beam training method is simply choosing the received signal with the highest energy and find its corresponding transmitting AWV. And, the state-of-the-art beamforming methods for mmWave MIMO systems are always derived with the assumption of full CSI, which brings high latency and impractical especially in the beginning phase of beamforming when CSI is completely unknown. However, this paper proposes a beamforming method in physical layer which aligns with the MAC-layer protocol IEEE 802.11 ay. The energy of the received signal as the coarse estimation channel is utilized to yield the optimal transmitter AWV with orthogonal codebook design and time complexity of $\mathcal{O}(n^2)$. This method is shown in the simulation to be of much higher capacity gain compared to 11 ad naive beam training method.

\bibliographystyle{IEEEtran}
\bibliography{ref}
\end{document}